\documentstyle[twocolumn,aps,epsf,rotate]{revtex}
\begin{document}
\twocolumn[\hsize\textwidth\columnwidth\hsize\csname
@twocolumnfalse\endcsname
 
\title{Unifying the Phase Diagrams of the Magnetic and Transport Properties of
La$_{2-x}$Sr$_x$CuO$_4$, $0 \leq x \lesssim 0.05$}
 
\author{E. Lai and R. J. Gooding}
 
\address{Department of Physics, Queen's University,\\ Kingston, Ontario K7L 3N6,
 Canada}
 
\date{\today}
\maketitle
 
\begin{abstract}
An extensive experimental and theoretical effort has led to a largely 
complete mapping of the magnetic phase diagram of La$_{2-x}$Sr$_x$CuO$_4$, and a 
microscopic model of the spin textures produced in the $x \lesssim 0.05$
regime has been shown to be in agreement with this phase diagram. Here we
use this {\em same} model to derive a theory of the impurity-dominated,
low temperature transport. Then, we present an analysis of previously
published data for two samples: $x = 0.002$ data from Chen {\em et. al.}, 
and $x = 0.04$ data from Keimer {\em et. al.} We show that the transport 
mechanisms in the two systems are the same, even though they
are on opposite sides of the observed insulator-to-metal transition.  
Our model of impurity effects on the impurity band conduction, 
variable-range hopping conduction, and coulomb gap conduction, is similar
to that used to describe doped semiconductors. However, for La$_{2-x}$Sr$_x$CuO$_4$ we find
that in addition to impurity-generated disorder effects, strong correlations
are important and must be treated on a equal level with disorder. On the 
basis of this work we propose a phase diagram that is consistent
with all available magnetic and transport experiments, and which connects 
the undoped parent compound with the lowest $x$ value for which La$_{2-x}$Sr$_x$CuO$_4$ 
is found to be superconducting, {\em viz.} $x \approx 0.06$.
 
\end{abstract}
 
\pacs{75.10.Jm,75.30.Ds}
]
\vskip 0.5cm
\narrowtext

\section{Introduction:}

The transport properties of La$_{2-x}$Sr$_x$CuO$_4$ at low temperatures have attracted 
considerable attention recently, in particular because it seems that 
for underdoped, superconducting Sr levels 
($0.06 \lesssim x \lesssim 0.15$) the normal state (superconductivity
suppressed by the application of a magnetic field) might be insulating
\cite {andobo,cieplak}. In this paper we present a theory for the
transport properties of La$_{2-x}$Sr$_x$CuO$_4$ for $x \lesssim 0.05$ with the 
expectation that one can better understand the superconducting
compounds if one first understands the weakly and moderately
doped non-superconducting materials. Our theory relies on treating
the effects of strong correlations and disorder with equal importance.

We employ a simple model to explain the low-temperature
transport (resistivity and magnetoresistance) of La$_{2-x}$Sr$_x$CuO$_4$ 
for $0.0 \leq x \lesssim 0.05$, and examine the evolution 
that occurs as the system is doped from the antiferromagnetic
insulator regime to the spin-glass phase. We stress that this
model is not a new invention contrived just to explain this data.
Instead, this same model has proven
successful in describing quantitatively the magnetic ``spin texture"
of this system \cite {goodingSG,salemTSF} for $0.0 \leq x \lesssim 0.05$. 
If any model is indeed a physically
realistic representation of La$_{2-x}$Sr$_x$CuO$_4$ it should be able to explain all of the
physics of this material, not just
the magnetic or transport properties. Thus, our present work on the
application of this same model to the transport behaviour in 
La$_{2-x}$Sr$_x$CuO$_4$ can be viewed as a critical test of the model. We find both
qualitative and quantitative agreement between this model and published data.
 
\section{Modelling of the Transport Data:}
\subsection{Approximations of our Transport Model}

To begin, let us clearly spell out the approximations implicit in our 
transport model. Assuming that the Sr impurities pin the carriers at low 
doping and low temperatures, it is now well established 
(see \cite{riceSK,goodingSK,rabeSK})
that the ground state corresponds to carriers circulating
either clockwise or counter-clockwise around the impurity.
Thus, one possible way to treat the coupling of the hole motion to the magnetic
background is to realize that the presence of strong correlations changes the
ground state from that of a circularly symmetric, $s$-wave impurity ground 
state, as is usually assumed, {\em e.g.}, for doped semiconductors,
to that of a doubly degenerate state with chiral quantum number
$\omega = \pm i$ \cite{goodingSK}; we refer to this as a chiral impurity ground
state.  As mentioned above, this model has been exploited in quantitatively
explaining a variety of experiments \cite {chouTSF,keimerCL,chouSG,goodingSG} 
concerning the magnetic properties of La$_{2-x}$Sr$_x$CuO$_4$.
In order to make progress on the modelling of the transport data we proceed as
follows.
 
A hole in the chiral impurity ground state circulates either clockwise or
counter-clockwise around a plaquette on the CuO$_2$ plane.
For our transport analysis we determine a continuum approximation for the 
chiral impurity ground state wave function by examining the 
Schr\"odinger equation in the effective mass approximation. We assume that
the hole is confined to the plane, so $\psi(x,y) = \psi(r,\phi)$. Then,
\begin{equation}
- \frac{{\hbar}^2}{2m^\ast}  \nabla^2 \psi -
\frac{e^2}{\epsilon_{\perp} \sqrt{r^2+d_\perp^2}} \psi = E \psi,
\label{eq:ShroEq}
\end{equation}
where the potential follows from the location of the impurities above or 
below each CuO$_2$ plane, and the out-of-plane dielectric constant, 
$\epsilon_{\perp}$.
For La$_{2-x}$Sr$_x$CuO$_4$ the appropriate numbers are $m^\ast \approx 1 - 2$ (we use 1.5),
$d_\perp \approx 1.85 \AA,$ and $\epsilon_\perp \approx 31$ \cite {chenprb2}.
(We note that experiment has shown \cite {dielectric} that $\epsilon_\perp$
is effectively independent of doping in the range $0 \leq x \lesssim 0.02$.)
Mimicking the circulating character of the ground state we determine
the wave functions of the form $\exp ({\pm i \phi})$. 

To compute the radial component of the wave function we note that
for $\omega = \pm i$ states asymptotically far from the impurity
the Schr\"odinger equation reduces to a form whose (radial) solutions can be
expressed in terms of Kummer's function. Then, using the asymptotic
properties of these functions, one finds that the continuum approximation
to the impurity wave function are	
\begin{equation}
\psi (r,\phi)~\sim r e^{-r/a} \ e^{\pm i \phi}
\label{eq:p-wave}
\end{equation}
where $a=({\epsilon_{\perp} \hbar^2}/{2 m^{\ast} e^2})$. Using the numbers
for La$_{2-x}$Sr$_x$CuO$_4$ given above, one has $a~\approx 5.48~\AA$. Of course, one
may also solve the radial Schr\"odinger equation for this problem
numerically, and in what follows all quantitative results are derived
from this more precise determination of the impurity wave function.
 
From now on we assume that transport proceeds by holes moving between
different impurity states, and ignore the effect of strong correlations
on the inter-impurity transit. That is,
we include the effects of strong correlations only by their influence on 
the specification of the {\em symmetry} of the impurity ground states. 
If the hole motion is between distant sites (as in Mott variable-ranged
hopping), or between neighbouring impurity sites (as in thermally activated,
impurity-band conduction), we simply use the numerical solution of 
Eq.~(\ref{eq:ShroEq}) with chiral symmetry
to predict the transport behaviour of La$_{2-x}$Sr$_x$CuO$_4$.
 
\subsection{Derivations of Conductivity for Different Temperature Regimes:}

Using the formalism of the Miller and Abrahams random resistor network 
model \cite{ES}, one can compute the transition probability 
$\langle \gamma_{ij} \rangle$ 
between any two impurity sites, $i$ and $j$. In what follows we will
present the resulting formulae for both the $s$-wave impurity
ground states, such that the relation with traditional doped semiconductor
work is clear, and for our chiral impurity ground states.

For $s$-wave symmetry impurity states, one finds
\begin{equation}
\langle \gamma_{ij}^{s} \rangle
\sim r_{ij}^{2d-4} \exp \left (-\frac{2r_{ij}}{a}-\frac{\epsilon_{ij}}{k_B T}\right)
\label{eq:s-wavecon}
\end{equation}
where $d=2,3$ is the dimensionality, $r_{ij}$ is the distance between the 
two sites, and $\epsilon_{ij}$ is the difference in on-site energies between 
the two sites. The activated form is well known, but the prefactor may not be;
in fact, usually the prefactor is ignored. However, this dependence
has appeared in the literature previously 
(see, e.g., Ref. \cite {ES}, Eqs. (4.2.17,18)).

The corresponding result for the chiral impurity ground states is
\begin{equation}
\langle \gamma_{ij}^{\chi} \rangle
\sim r_{ij}^{2d} \exp \left (-\frac{2r_{ij}}{a}-\frac{\epsilon_{ij}}{k_B T}\right ).
\label{eq:p-wavecon}
\end{equation}
Note that the only change is in the $r$ dependence of the prefactor.

Applying percolation theory \cite{ES} to 
Eqs.~(\ref{eq:s-wavecon},\ref{eq:p-wavecon}), 
one may derive the conductivity for these theories as a function
of temperature, and three different regimes are found. The transport
is always phonon-assisted in that energy must be supplied to 
(absorbed from) a hole localized on site $i$ in order that it can
move to site $j$ (assuming that site $j$ is unoccupied). 
At high temperatures, there are phonons of all energies available, so
that the hole can always hop to its neighbouring impurity site, regardless
of the difference in their on-site energies. Hence, 
nearest neighbour, or so-called impurity band conduction (IBC), takes place.
The conductivity is then given by
\begin{equation}
\sigma_{IBC}(T) \sim \exp \: \left(-\frac{\epsilon_c}{k_B T} \right)
\label{eq:IBC}
\end{equation}
where $\epsilon_c$ is the average activation energy needed for a carrier to
move to its neighbouring site. This result is independent of dimensionality 
and the symmetry of the impurity ground state.

As the temperature is lowered, motion between neighbouring sites may be 
forbidden due to the lack of phonons of appropriate energy. Consequently, 
it is more 
likely for the carriers to hop to a more distant site if this means that the 
energy difference is less. This is known as Mott variable range hopping (VRH). 
The conductivity for $d$-dimensional variable range hopping for conventional
$s$-wave impurities is given by the familiar expression
\begin{equation}
\sigma_{VRH,s}(T) \sim
\left(\frac{1}{T} \right)^{\frac{2d-4}{d+1}} \, \exp \: \left(- \left(
\frac{T_{d}}{T} \right)^{1/(d+1)}
\right)
\label{eq:VRH,s}
\end{equation}
where $T_d$ is a characteristic temperature given by
\begin{eqnarray}
T_3 & = & \frac{22.8}{g(\mu) k_B a^3} \\
T_2 & = & \frac{13.8}{g(\mu) k_B a^2}
\label{eq:chartemp}
\end{eqnarray}
for three and two dimensions, respectively.  Here $g(\mu)$ is the density
of states at the Fermi level, which is assumed to be constant in the 
VRH regime \cite{ES}. Ignoring the prefactor, as is usually done \cite {ES},
for 3D systems one has the familiar ``Mott 1/4 law" for VRH.
For the chiral impurity ground states that we are considering, one finds
\begin{equation}
\sigma_{VRH,\chi}(T) \sim
\left(\frac{1}{T} \right)^{\frac{2d}{d+1}} \, \exp \: \left(- \left(
\frac{T_{d}}{T} \right)^{1/(d+1)}
\right)~~,
\label{eq:VRH,p}
\end{equation} 
and thus only the temperature dependence of the prefactor is different.

For temperatures so low such that the energy difference between the initial
and final site is comparable to the Coulomb correlation energy between 
carriers, the density
of states near the Fermi level is no longer constant.  Instead, Coulomb
interactions cause the density of states to vanish at the Fermi level \cite{ES},
and a Coulomb gap is formed.  
A model known as the Coulomb gap model (CG) can be used to describe the conductivity in
this situation, and the result is 
\begin{equation}
\sigma_{CG,s}(T) \sim \left( \frac{1}{T} \right)^{d-2}\,
\exp\: \left(-{\left(\frac{T_{ES}}{T} \right)}^{\frac{1}{2}} \right)
\label{eq:CG,s}
\end{equation}
for $s$-wave states,
where
\begin{equation}
T_{ES} = \frac{2.9\,e^2}{\epsilon_{\parallel} a k_B}
\label{eq:EStemp}
\end{equation}
and $\epsilon_{\parallel}$ is the in-plane dielectric constant.
The corresponding result for chiral impurity states is
\begin{equation}
\sigma_{CG,\chi}(T) \sim \left( \frac{1}{T} \right)^{d}\,
\exp\: \left(-{\left(\frac{T_{ES}}{T} \right)}^{\frac{1}{2}} \right)~~.
\label{eq:CG,p}
\end{equation}
The exponent $1/2$ inside the exponential factor is the same for both 
two and three dimensions, but again the prefactors depend on the symmetry
of the impurity ground state.
 
This sequence of impurity-dominated transport as a function of
temperature, {\em viz.} IBC at high temperatures, VRH at intermediate 
temperatures, followed by CG transport at low temperatures, has
been observed in a number of experimental systems.
For example, in amorphous Ge IBC has been observed for 
$T \gtrsim 200K$, VRH conduction occurs
for $2K \lesssim T \lesssim 200K$ \cite{ge1,ge2,ge3}, and
CG conduction was observed for $T \lesssim 2K$ \cite{ge1,ge2,ge3}.  
However, the prefactors that we have obtained were not included
in these analyses. In what follows we use these terms to determine
the impurity ground state symmetry, and find strong support
for the chiral impurity ground state model used previously to describe
successfully the magnetic properties of La$_{2-x}$Sr$_x$CuO$_4$.
 
\subsection{Analysis of Previously Published Conductivity Data}

In the previous section we presented our predictions for the
temperature dependence of the conductivity due to various kinds 
of hopping conduction processes in different temperature regimes.  
The expressions for these conductivities are summarized in Table I.
In order to fit these expressions to data, we note that for all theories
there are two fitting parameters. Thus, minimizing the $\chi^2$ of a fit
is equivalent to maximizing the goodness of fit parameter, 
and in what follows we only refer to the $\chi^2$.
\begin{table}
\narrowtext
\caption{Summary of the differing temperature dependencies of the conductivity.}
\label{tab:conds}
\begin{center}
\begin{tabular}{|c|c|}
IBC & $\sigma_{IBC} (T) \propto \exp \left(-{\epsilon_c}/{k_B T} \right)$ \\
\hline \hline
3-D $s$-wave VRH &
$\sigma_{VRH,s}(T)  \propto T^{-{2}/{3}}
\exp \left( -\left({T_{3}}/{T} \right)^{{1}/{4}} \right)$ \\ \hline
2-D $s$-wave VRH &
$\sigma_{VRH,s}(T)  \propto
\exp \left( -\left({T_{2}}/{T} \right)^{{1}/{3}} \right)$ \\ \hline
3-D chiral VRH &
$\sigma_{VRH,\chi}(T)  \propto T^{-{3}/{2}}
\exp \left( -\left({T_{3}}/{T} \right)^{{1}/{4}} \right)$ \\ \hline
2-D chiral VRH &
$\sigma_{VRH,\chi}(T)  \propto T^{-{4}/{3}}
\exp \left( -\left({T_{2}}/{T} \right)^{{1}/{3}} \right)$ \\ \hline \hline
3-D $s$-wave CG &
$\sigma_{CG,s}(T) \propto T^{-1}
\exp \left(-{\left({T_{ES}}/{T} \right)}^{{1}/{2}} \right)$ \\ \hline
2-D $s$-wave CG &
$\sigma_{CG,s}(T) \propto
\exp \left(-{\left({T_{ES}}/{T} \right)}^{{1}/{2}} \right)$ \\ \hline
3-D chiral CG &
$\sigma_{CG,\chi}(T) \propto T^{-3}
\exp \left(-{\left({T_{ES}}/{T} \right)}^{{1}/{2}} \right)$ \\ \hline
2-D chiral CG &
$\sigma_{CG,\chi}(T) \propto T^{-2}
\exp \left(-{\left({T_{ES}}/{T} \right)}^{{1}/{2}} \right)$ \\
\end{tabular}
\end{center}
\end{table}
The first data that we analyzed was that of a $x=0.002$ single crystal prepared 
by Chen {\em et al.} \cite{chenprb2}. Their analysis ignored the 
temperature prefactors mentioned above, and led to the conclusion that
the transport in this weakly doped antiferromagnet could be described by
conventional doped semiconductor theory. In particular, IBC was observed for
$50K \lesssim T \lesssim 295K$, while for $4K \lesssim T \lesssim 50K$ they 
found that the conduction mechanism is that of the traditional 3-D Mott
VRH type. No data below $4 K$ was taken, and thus no evidence of CG conduction
was found.

We agree with their interpretation of IBC for $50 K \lesssim T \lesssim 295 K$, 
and from our fit we find $\epsilon_c \approx 0.020eV$ as the average 
activation energy needed for the hop. However, 
for $4 K \lesssim T \lesssim 50 K$, we find that the chiral VRH expression
from Eq.~(\ref{eq:VRH,p}) with $d=2$ gives a better fit to the conductivity 
data. A comparison of all four VRH hopping theories mentioned above
is provided in Table II. 
 
We also studied the conductivity data of a $x=0.04$ single crystal prepared 
by Keimer {\em et al.} \cite{keimerCL}. For reasons that are unclear to
us, around $50 K$ only one conductivity theory was compared to
the data, that being appropriate for a system displaying 2-D weak
localization. For such a system one predicts a logarithmic dependence of
temperatures (which also has two fitting parameters). 
\begin{table}
\narrowtext
\caption{Comparison of VRH theories for the $x = 0.002$ Chen data,
for $50 K \leq T \leq 295 K$.}
\label{tab:vrh}
\begin{center}
\begin{tabular}{|c|c|c|c|c|}
 & 3-D $s$-wave & 2-D $s$-wave & 3-D chiral & 2-D chiral \\ \hline
$\chi^2$ & 0.411 & 0.371 & 0.317 & 0.169 \\ \hline
$T_d$ & $5.1 \times 10^6 K$ & $6.7 \times 10^4 K$ & $9.3 \times 10^6 K$ &
$1.3 \times 10^5 K$ \\
\end{tabular}
\end{center}
\end{table}
\noindent
These authors
suggested that such a conduction mechanism was operative from 
$10 K \lesssim T \lesssim 100 K$. Then, for $1 K \lesssim T \lesssim 10K$
the system was in a ``crossover" regime, and then for $T \lesssim 1 K$
the system displayed CG conduction. Clearly, if such a description
were true, except at the lowest temperatures the conduction mechanisms
of the $x = 0.002$ and the $x = 0.04$ systems would have nothing to
do with one another.

In contrast to this approach, we disagree with using weak localization theory
to account for the conduction mechanism in La$_{2-x}$Sr$_x$CuO$_4$ around $50 K$.
Firstly and most importantly, weak localization theory fails to include
the strong correlation effects between the hole
and the background Cu spins, in direct contrast to what is made manifest by
studies of the magnetic properties of this material.  
Secondly, such behaviour is completely at odds with the negative,
{\em isotropic} magnetoresistance observed by this same group \cite {preyer}.
After reanalyzing their conductivity data, we find a much simpler and more
natural explanation --- for $20K \lesssim T \lesssim 70K$, Eq.~(\ref{eq:IBC}), 
the simple activated expression for IBC, gives a 
much better fit than the logarithmic 
temperature dependence arising from weak localization theory. The $\chi^2$ for
these fits is listed in Table III, and it is seen that IBC has a $\chi^2$
at least 20 times less than the corresponding $\chi^2$ for
logarithmic temperature dependence.  The average activation
\begin{table}
\narrowtext
\caption{Comparison of the IBC and the weak localization predictions for the
$x = 0.04$ Keimer data, for $20 K \leq T \leq 70 K$.}
\begin{center}
\begin{tabular}{|c|c|c|}
 & IBC~~~~~~~~~~~~ & weak localization ~~~~~~\\ \hline
$\chi^2$ & 0.000618~~~~~~~~~~~~ & 0.0142 ~~~~~~~~\\
\end{tabular}
\end{center}
\end{table}
\begin{table}
\narrowtext
\caption{Comparison of VRH theories for the $x = 0.04$ Keimer data,
for $1 K \leq T \leq 20 K$.}
\label{tab:vrh1}
\begin{center}
\begin{tabular}{|c|c|c|c|c|}
 & 3-D $s$-wave & 2-D $s$-wave & 3-D chiral & 2-D chiral \\ \hline
$\chi^2$ & 0.0756 & 0.0795 & 0.0823 & 0.0483 \\ \hline
$T_d$ & $2.7 \times 10^4 K$ & $550 K$ & $1.2 \times 10^5 K$ &
$630 K$ \\
\end{tabular}
\end{center}
\end{table}
\begin{table}
\narrowtext
\caption{Comparison of CG theories for the $x = 0.04$ Keimer data, for
$T < 1 K$.}
\begin{center}
\begin{tabular}{|c|c|c|c|c|}
& 3-D $s$-wave & 2-D $s$-wave & 3-D chiral & 2-D chiral \\ \hline
$\chi^2$ & 0.291 & 0.286 & 1.633 & 0.271 \\ \hline
$T_{ES}$ & $30 K$ & $51 K$ & $80 K$ & $32 K$ \\
\end{tabular}
\end{center}
\end{table}
\noindent
energy found from the fit is $\epsilon_c \approx 0.002eV$, roughly a factor
of ten less than that for the $x = 0.002$ sample. The poor comparison
of weak localization theory is not improved when a different temperature
regime is used.
 
If our chiral impurity model is indeed correct, and weak localization is not
found for this system, it should exhibit a crossover from IBC to VRH
as the temperature is lowered, and this is precisely what we find.  
For $1K \lesssim T \lesssim 20K$, we find that Eq.~(\ref{eq:VRH,p}) 
with $d=2$, the expression for 2-D chiral VRH again gives the
best fit to the conductivity data --- the statistical data is listed in
Table IV.
 
Below 1K we find that Eq.~(\ref{eq:CG,p}) with $d=2$,
the expression for 2-D chiral impurity Coulomb gap
hopping gives the best fit to the conductivity data; Table V summarizes
the statistical data. 

In {\em all} cases discussed above, the best fit to the data is found to 
correspond to our 2-D chiral impurity ground state theory. 
We believe that this {\em repeated agreement} between
theory and experiment lends strong credibility to our model, an argument
that is further strengthened when it is noted that this same model
successfully describes the magnetic properties of La$_{2-x}$Sr$_x$CuO$_4$.
 
\subsection{Crossover Temperatures}

In the previous section we showed that our 2-D chiral conduction model, 
which follows from the chiral impurity ground state generated by
strong correlations, 
fits the experimental data better than other available theories. Further
quantitative support for this theory follows from a study of the
crossover temperatures, which we now present.

For the $x=0.002$ crystal there is a crossover from IBC to 2-D chiral VRH
conduction at around 50K.  For the $x=0.04$ crystal,
the crossover from IBC to 2-D chiral VRH conduction
occurs at around 20K.  Then, the crossover to 2-D chiral CG conduction
occurs at around 1K. 

One may determine these crossover temperatures semi-empirically.
That is, in what follows we derive the crossover temperatures theoretically,
but our formulae involve parameters which we cannot determine. However, we can 
express these parameters in terms of the activation energy $\epsilon_c$, 
and the characteristic temperatures $T_2$ and $T_{ES}$.
All of these numbers were determined in the previous
section from the fits of our model to the experimental data, and are
stated in the text and/or listed in the tables. 
 
The crossover between IBC and VRH can be estimated from the condition that 
the average activation energy between neighbouring sites is equal to 
the activation energy for carriers executing variable range hopping.
The latter quantity can be determined from expressing the hopping distance
between two sites in terms of the activation energy, and then setting
the hopping distance to be the average distance as a function of
temperature. That is, the activation energy for carriers executing variable 
range hopping is given by $- {d \, {\rm ln} \sigma_{VRH}}/{d \beta}$ \cite{ES}, 
where $\beta \equiv {1}/{k_B T}$, and  $\sigma_{VRH}$ is given by the
the chiral expression in Eq.~(\ref{eq:VRH,p}) with $d=2$.
Hence, by solving
\begin{equation}
\epsilon_c = - {d~{\rm ln}~(\sigma_{VRH})\over d~\beta},
\label{eq:txvrh}
\end{equation}
one obtains
\begin{equation}
T_{IBC \rightarrow VRH} = \frac{1}{\sqrt{T_2}}
\left(\frac{3 \epsilon_c}{k_B} \right)^{\frac{3}{2}}.
\label{eq:crossIBCtoVRH}
\end{equation}

For the $x=0.002$ crystal, we take $T_2 = 1.3 \times 10^5 K$ from
Table II, and our fit to IBC yielded $\epsilon_c = 0.020 eV$.
This produces $T_{IBC \rightarrow VRH} = 51K$, in good agreement
with the experimental value of $50 K$ \cite{chenprb2}.
For the $x=0.04$ crystal we take $T_2 = 630 K$ from Table IV, and
our fit yeilded $\epsilon_c = 0.002 eV$.  
This gives $T_{IBC \rightarrow VRH} = 23 K$, again in good agreement
with the experimental value of $20 K$ \cite{keimerCL}.
 
One may derive the crossover temperature from VRH to
CG conduction as well.  As discussed previously, this crossover occurs when
the Coulomb correlation energy between carriers is equal to the activation 
energy needed to hop to a distant site.  The former is given by
${e^2}/{\epsilon_{\parallel} \bar{r}}$, where $\bar{r}$ is the most likely
hopping distance for variable range hopping. Hence by solving
\begin{equation}
{e^2}/{\epsilon_{\parallel} \bar{r}}= - {d~{\rm ln}~(\sigma_{VRH})\over d~\beta}
\label{eq:txcg}
\end{equation}
one obtains
\begin{equation}
T_{VRH \rightarrow CG} = 33.2 \frac{T_{ES}^3}{T_{2}^{2}}
\label{eq:crossVRHtoCG}
\end{equation}
as the crossover temperature between variable range hopping and Coulomb gap
conduction.

For the $x=0.04$ crystal, $T_2 = 630 K$ from Table IV and $T_{ES} = 32K$ from
Table V. This gives $T_{VRH \rightarrow CG} \approx 2K$, again in good 
agreement with the experimental value of $1 K$ \cite{keimerCL}.

The other theories discussed in the previous sections can also
be used to calculate crossover temperatures, but none of these sets of
temperatures accurately tracks the experimental crossover temperatures
as well as our 2-D chiral theory. Thus, this set of crossover temperatures 
provides further support quantitatively for our model.
 
\section{Metal-to-Nonmetal Transition Doping Concentration}

La$_{2-x}$Sr$_x$CuO$_4$ is a strongly correlated electronic system which is greatly
influenced by the disorder effects produced by the Sr impurities.
This view is made clear by the above transport work and previously
published work on the magnetic properties \cite {goodingSG}.
It is known that this system undergoes a nonmetal-to-metal transition
with increasing doping level $x$, and in this section we employ
our chiral impurity model to show that it also reproduces the
experimental value for the critical doping level, 
$x_c \approx 0.02$. 
However, before proceeding to this derivation, we wish to present a 
clear discussion of how we believe this transition should be viewed, 
since confusing and conflicting remarks dominate the literature on 
this subject.

The conventional view of a nonmetal-to-metal transition in a disordered
system has been demonstrated for many experimental systems; here we use 
the recent, elegant work of Dubon, {\em et al.} \cite {dubonprl}, for Ge 
(under stress) doped with Cu impurities to demonstrate this.
Upper and lower Hubbard impurity ``bands" form, and the
gap separating these bands decreases with increasing impurity concentration.
The critical concentration is that at which this gap closes. However,
for Cu doping levels beyond the critical concentration, the transport
at low temperatures is still found to be insulating like, {\em viz.}
the resistivity increases with decreasing temperature \cite {dubon} 
--- only at higher temperatures is the chemical potential pushed
through the mobility edge and metallic conduction is found. This 
phenomenology thus implies that the temperature at which metallic
conduction is found (by which we imply the resistivity increases with 
increasing temperature) is concentration dependent, and that the
critical concentration for the nonmetal-to-metal transition is the
doping level for which metallic conduction is first found at
high temperatures.

This picture is consistent with published transport data 
for La$_{2-x}$Sr$_x$CuO$_4$ at low and intermediate doping levels:
At the lowest Sr levels the resistivity is always found to be
insulating like \cite {chenprb2,datalt}, while for larger 
$x$ \cite {datagt} only at high temperatures does the resistivity 
increase (approximately linearly) with increasing temperature.
This is further substantiated from the dielectric constant measurements of 
Chen, {\em et al.} \cite {dielectric} who found that the in-plane dielectric 
constant of La$_{2-x}$Sr$_x$CuO$_4$ saturated at high frequencies,
and this saturation value diverged at some doping level. It is difficult
to infer the critical $x$ value from this experiment, and in what
follows we use $x_c \approx 0.02 \pm 0.005$ as a reasonable approximation 
for the critical concentration of the metal-to-nonmetal transition.

Our chiral impurity model can be used to estimate the critical doping
concentration for La$_{2-x}$Sr$_x$CuO$_4$ via the Mott-Hubbard treatment 
of such transitions. 
The overlap integral between impurity sites separated by a distance $r$,
and the on-site Coulomb repulsion for two holes being at one impurity
site, can be estimated using the wave function given in Eq.~(\ref{eq:p-wave}).
Then, the Mott criterion corresponds to the doping level at which
an impurity ``band" has a width equal to the on-site Coulomb
repulsion energy. These simple calculations \cite {zeff} lead to the 
2-D analogue of Mott's criterion for chiral impurity ground states, 
{\em viz.} 
\begin{equation}
{x_c}^{\frac{1}{2}}\, \left(\frac{a}{a_0} \right) \approx 0.191
\label{eq:MNM}
\end{equation}
where $a$ is the effective Bohr radius of the chiral impurity ground
state (stated previously to be 5.48 \AA), and $a_0=3.85 \AA$ is the planar
lattice constant.  Solving this equation we find $x_c \approx 0.018$, 
in excellent agreement with experiment. 
The fact that our theory is {\em two} dimensional agrees also with the 
experimental fact that even though the
in-plane dielectric constant at high frequencies diverges at $x_c$, the
out-of-plane dielectric constant remains roughly constant
as $x_c$ is approached \cite {dielectric}.
Thus, quantitatively and qualitatively, we find
that our chiral impurity model is consistent with available, published
transport data on the metal-to-nonmetal transition.

\section{$\rho_{ab}$ phase diagram}

Figure 1 shows the approximate $\rho_{ab}$ transport phase diagram 
for the new hopping conduction mechanism that we propose for 
$x \lesssim 0.05$, and summarizes our results.  
From this figure it is clear that our work supports the
contention that at low temperatures the conduction mechanism on 
either side of the metal-to-nonmetal transition ($x_c \approx 0.02$) 
is the same. This is in disagreement with the idea proposed in
Ref. \cite {keimerCL} that weak localization effects are seen
in the $x=0.04$ transport data.  
For $x \gtrsim 0.02$ and at high $T$, La$_{2-x}$Sr$_x$CuO$_4$ behaves like
an anomalous metal with an approximate $T$-linear resistivity. Our theory
has nothing to say about the dominant scattering mechanism that
produces this unusual behaviour.
 
In Fig.~1 the superconducting phase for $x \gtrsim 0.05$ and $T<T_c$ is 
shown. The region between the superconducting phase and the anomalous 
metallic phase is referred to as an ``anomalous insulator'' phase according 
to Ando {\em et al.} \cite{andobo}. They examined the normal state properties 
of La$_{2-x}$Sr$_x$CuO$_4$ ($x=0.08$ and $x=0.013$) down to 
$T/T_c \approx 0.04$ by suppressing 
superconductivity with a pulsed magnetic field of 61T along the $c$-axis.  
They measured the in-plane resistivity $\rho_{ab}$ and the
out-of-plane resistivity $\rho_c$, and found insulating behaviour for both 
resistivities at low temperatures. Under the assumption that the magnetic 
field dependence of the resistivity is very small compared to the temperature 
dependence, Ando {\em et al.} claim that their materials are indeed insulating 
in the region labelled ``anomalous insulator'' in Fig.~1. 
(However, their main assumption has been challenged by the work of Malinowski 
{\em et al.} \cite{cieplak}, who claimed that the behaviour
observed in strong magnetic fields is not a reliable guide to the nature 
of the zero-field ground state in the absence of superconductivity.  
Malinowski {\em et al.} measured the normal state conductances per 
CuO$_2$ plane for two highly underdoped
superconducting La$_{2-x}$Sr$_x$CuO$_4$ samples ($x=0.048$ and $x=0.051$) at different fields,
and their data is found to collapse onto one curve with the use of a single
scaling parameter that is inversely proportional to the Bohr radius of the 
ground state wave function. When extrapolated to zero field, this scaling 
parameter approaches zero, which suggests that the zero-field ground state 
may be extended, as opposed to localized (as suggested by Ando {\em et al.}). 
This discrepancy between the two groups has not yet been settled.)
 
The relationship of our work to this portion of the phase diagram is
unclear. If disorder effects are important, then it is certainly possible
that the anomalous insulating behaviour (if truly present) might be related 
to the physics discussed in this paper.
Alternatively, as proposed in Ref. \cite {cieplak}, if
the anomalous insulating behaviour is associated with the magnetic
field changing the electronic structure and subsequently producing
localized states, then it is unlikely that our theory can be extrapolated
to such doping levels.

Further, we hope that new data on travelling-solvent
float-zone grown single crystals for a variety of doping levels
above and below $x = 0.05$ \cite {japan} will allow for us to judge 
conclusively if
it is appropriate to extrapolate our theory to the doping concentrations 
containing the superconducting samples.
\begin{figure}[h]
\setlength{\unitlength}{1in}
\begin{picture}(0,4)
\epsfxsize=8cm
\put(-.2,-.2){\epsffile{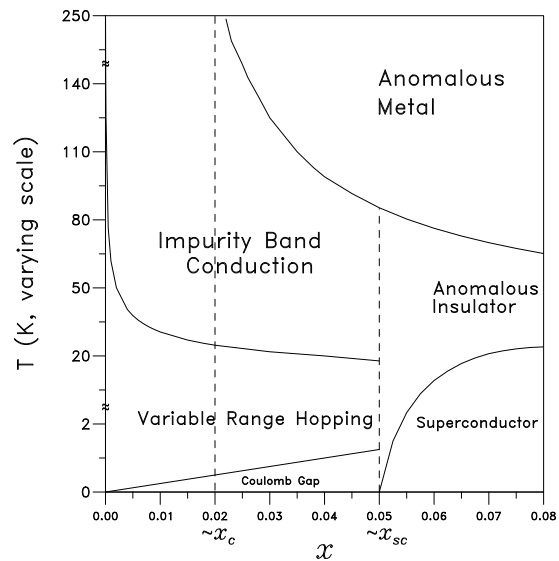}}
\end{picture}
\narrowtext
\caption{
$\rho_{ab}(T)$ phase diagram for La$_{2-x}$Sr$_x$CuO$_4$ summarizing its transport
properties as inferred from our analysis. The dashed line at
$x_c \approx 0.02$ is the critical doping concentration for the
metal-to-nonmetal transition.  The dashed line at $x_{sc} \approx 0.05$
is the onset for superconductivity --- this vertical line on our phase
diagram is a guide for the eye only. The anomalous metallic phase at
high temperatures for $x > 0.02$ and the superconducting phase for
$x > 0.05$ are also shown in this figure.
The nature and the extent of the ``anomalous insulator'' part of the
phase diagram is not yet known.
}
\label{fig1}
\end{figure}

\section{Summary:}
 
To summarize, we have presented a theory of
the transport of La$_{2-x}$Sr$_x$CuO$_4$ for $x \lesssim 0.05$.
It is to be emphasized that the physics that led to our theory,
the chiral impurity ground state, and the successful description
of the spin texture of La$_{2-x}$Sr$_x$CuO$_4$ at low temperatures \cite {goodingSG},
are the same. Disorder and strong correlations dominate the low
temperature, low doping regime of La$_{2-x}$Sr$_x$CuO$_4$.

One potential weakness of our work is that almost all of our comparisons 
are of a quantitative nature. Instead, one would like to compare the 
qualitative behaviour found in certain transport measurements. To this end,
we are presently preparing a manuscript on a comparison of our
theory to the magnetoresistance measurements for samples in this
same doping regime. In particular, such work allows for the
scaling properties of a theory, with respect to field and temperature,
to be compared to experiment.  This theory is outside of the simple treatment
of impurity-dominated hopping-type conduction given here, and will
be reported elsewhere.

\vspace*{-0.5cm}
\acknowledgments

We wish especially to thank Chih-Yung Chen and Bernhard Keimer for providing
us with listings of their previously published data, 
and Shiki Ueki, Kazu Yamada, Yasuo Endoh, Lance Miller,
Oscar Dubon and Eugene Haller for providing us with unpublished data. 
One of us (RJG) wishes to thank Marc Kastner for introducing him
to the MIT experimental work on this subject. We thank Yoichi Ando, 
Lance Miller, Tom Timusk, Bob Birgeneau and David Johnston for critical 
comments. This work was begun while one of us (RJG) was visiting
MIT and McMaster University, and he wishes to thank them for their
hospitality. This work was supported by the NSERC of Canada.


\vspace*{-0.5cm}
\begin{references}
\vspace*{-1.5cm}

\bibitem{andobo} Y. Ando, G. S. Boebinger, A. Passner, Tsuyoshi Kimura, 
Kohji Kishio, Phys. Rev. Lett.
{\bf 75}, 4662 (1995); Y. Ando, G. S. Boebinger, A. Passner,
N. L. Wang, C. Geibel, F. Steglich,  Phys. Rev. Lett. {\bf 77}, 2065 (1996); 
G. S. Boebinger, Y. Ando, A. Passner,
T. Kimura, M. Okuya, J. Shimoyama, K. Kishio,
K. Tamasaku, N. Ichikawa, S. Uchida, Phys. Rev. Lett. 
{\bf 77}, 5417 (1996); Y. Ando, G. S. Boebinger, A. Passner, Tsuyoshi Kimura,
Kohji Kishio, J. Low Temp. Phys. {\bf 105}, 867 (1996). 

\bibitem{cieplak} K. Karpinska, A. Malinowski, 
M. Z. Cieplak, S. Guga, S. Gershman, G. Kotliar,
T. Skoskiewicz, W. Plesiewicz, M. Berkowski, P. Lindenfeld, 
Phys. Rev. Lett. {\bf 77},
3033 (1996); A. Malinowski, M. Z. Cieplak, A. S. van Steenbergen, 
J. A. A. J. Perenboom, K. Karpinska, M. Berkowski, S. Guha, P. Lindenfeld, 
Phys. Rev. Lett. {\bf 79}, 495 (1997).
 
\bibitem{goodingSG} R. J. Gooding, N. M. Salem, R. J. Birgeneau, and 
F. C. Chou, Phys. Rev. B {\bf 55}, 6360 (1997). 

\bibitem{salemTSF} N. M. Salem, R. J. Gooding, and A. Mailhot, Phys. Rev.
B {\bf 49}, 6067 (1994).
 
\bibitem{riceSK} K. J. von Szczepanski, T. M. Rice, and F. C. Zhang, 
Europhys. Lett. {\bf 8}, 797 (1989).

\bibitem{goodingSK} R. J. Gooding, Phys. Rev. Lett. {\bf 66}, 2266 (1991).

\bibitem{rabeSK} K. M. Rabe, and R. Bhatt, J. Appl. Phys. {\bf 69}, 4508 (1991).
 
\bibitem{chouTSF} F. C. Chou, F. Borsa, J. H. Cho, D. C. Johnston, 
A. Lascialfari, D. R. Torgeson, and J. Ziolo, 
Phys. Rev. Lett. {\bf 71}, 2323 (1993). 
 
\bibitem{keimerCL} B. Keimer, N. Belk, R. J. Birgeneau, A. Cassanho, C. Y. Chen,
M. Greven, M. A. Kastner, A. Aharony, Y. Endoh, R. W. Erwin, and G. Shirane, 
Phys. Rev. B {\bf 46}, 14034 (1992).
 
\bibitem{chouSG} F. C. Chou, N. R. Belk, M. A. Kastner, R. J. Birgeneau, 
and A. Aharony, Phys. Rev. Lett. {\bf 75}, 2204 (1995).
 
\bibitem{chenprb2} C. Y. Chen, E. C. Branlund, ChinSung Bae, K. Yang,
                   M. A. Kastner, A. Cassanho, and R. J. Birgeneau, Phys. Rev.
                   B {\bf 51}, 3671 (1995).

\bibitem{dielectric}
C. Y. Chen, N. W. Preyer, P. J. Picone, M. A. Kastner, 
H. P. Jenssen, D. R. Gabbe, A. Cassanho, and R. J. Birgeneau, 
Phys. Rev. Lett. {\bf 63}, 2307 (1989); N. W. Preyer, R. J. Birgeneau,
C. Y. Chen, D. R. Gabbe, H. P. Jenssen, M. A. Kastner, P. J. Picone,
Tineke Thio, Phys. Rev. B {\bf 39}, 11563 (1989); C. Y. Chen, R. J. Birgeneau,
M. A. Kastner, N. W. Preyer, Tineke Thio, Phys. Rev. B {\bf 43}, 392 (1991).

\bibitem{ES} For a review, see \lq\lq Electronic Properties of Doped 
Semiconductors,'' B. I. Shklovskii and A. L. Efros 
(Springer-Verlag, New York, 1984).

\bibitem{ge1} N. F. Mott, Phil. Mag. {\bf 26}, 1015 (1972).

\bibitem{ge2} A. H. Clark, Phys. Rev. {\bf 154}, 750 (1967).

\bibitem{ge3} G. Sadasiv, Phys. Rev. {\bf 128}, 1131 (1962).

\bibitem{preyer} N. W. Preyer, M. A. Kastner, C. Y. Chen, R. J. Birgeneau, and
                     Y. Hidaka, Phys. Rev. B {\bf 44}, 407 (1991).

\bibitem{dubonprl} O. D. Dubon. W. Walukiewicz, J. W. Beeman,
and E. E. Haller, Phys. Rev. Lett. {\bf 78}, 3519 (1997).

\bibitem{dubon} O. Dubon (private communication).

\bibitem{datalt} One $\rho_{ab}(T)$ measurement for
La$_{1.99}$Sr$_{0.01}$CuO$_4$ has been taken up to room temperature:
A. Lacerda, T. Graf, J. H. Cho, J. D. Thompson, M. P. Maley,
Physica C {\bf 235-240}, 1353 (1994).
L. Miller (private communication) has measured the
in-plane resistivity of the related single-plane system
${\rm Sr_2 CuO_2 Cl_2}$, which has an approximate
doping level (in terms of the $x$ variable describing the Sr levels of La$_{2-x}
$Sr$_x$CuO$_4$)
of $x \approx 0.009$. His data is consistent with the activated form
of nearest-neighbour impurity band conduction, and was found to occur
for $200K \lesssim T \lesssim 350K$. Around $375K$, the sample reduced
in a reduced--oxygen atmosphere, and hence created problems for accurate
measurements at higher temperatures.

\bibitem{datagt} H. Takagi, B. Batlogg, H. L. Kao, J. Kwo, R. J. Cava,
J. J. Krajewski, W. F. Peck, Phys. Rev. Lett. {\bf 69}, 2975 (1992).

\bibitem{zeff} Implicit in this expression is the assumption that
the effective coordination number is equal to 2, the dimensionality
of the hopping conduction. This may be justified in general, as will
be discussed in a future publication, or on the basis of the
topology of the spin texture discussed in Ref. [1].

\bibitem{japan} S. Ueki, K. Yamada, and Y. Endoh (private communication).

\end{references}
\end{document}